\documentclass[nofootinbib,aps,prl,superscriptaddress,groupedaddress]{revtex4}  % for review and submission
\usepackage{graphicx}  % needed for figures
\usepackage{dcolumn}   % needed for some tables
\usepackage{bm}        % for math
\usepackage{amssymb}   % for math

\def\beq{\begin{eqnarray}}
\def\eeq{\end{eqnarray}}

\def\({\left(}
\def\){\right)}

\def\mpl{M_{\rm pl}}

\newcommand{\be}{\begin{equation}}
\newcommand{\ee}{\end{equation}}

\def\ea{\end{eqnarray}}
\def\ba{\begin{eqnarray}}

\def\beq{\begin{eqnarray}}
\def\eeq{\end{eqnarray}}

\def\({\left(}
\def\){\right)}

\def\mpl{M_{\rm pl}}

\def\lsim{\mathrel{\rlap{\lower3pt\hbox{\hskip0pt$\sim$}}
     \raise1pt\hbox{$<$}}}         %less than or approx. symbol
\def\gsim{\mathrel{\rlap{\lower4pt\hbox{\hskip1pt$\sim$}}
     \raise1pt\hbox{$>$}}}         %greater than or approx. symbol

\def\lsim{\mathrel{\rlap{\lower3pt\hbox{\hskip0pt$\sim$}}
     \raise1pt\hbox{$<$}}}         %less than or approx. symbol
\def\gsim{\mathrel{\rlap{\lower4pt\hbox{\hskip1pt$\sim$}}
     \raise1pt\hbox{$>$}}}         %greater than or approx. symbol

\usepackage{wrapfig}
\usepackage{amsmath}
\usepackage{amsfonts}
\usepackage{verbatim}
\usepackage{graphicx}
\usepackage{subfigure}
\usepackage{amssymb}
\usepackage[T1]{fontenc}
\usepackage{slashed}
\usepackage{color}
\usepackage{lipsum}
\usepackage[export]{adjustbox}
\begin{document}

\title{Core Fragmentation in Simplest Superfluid Dark Matter Scenario}
\author{Lasha Berezhiani}
\address{Max-Planck-Institut f\"ur Physik, F\"ohringer Ring 6, 80805 M\"unchen, Germany}
\address{Arnold Sommerfeld Center, Ludwig-Maximilians-Universit\"at, Theresienstra{\ss}e 37, 80333 M\"unchen, Germany}
\author{Giordano Cintia}
\address{Max-Planck-Institut f\"ur Physik, F\"ohringer Ring 6, 80805 M\"unchen, Germany}
\author{Max Warkentin}
\address{Arnold Sommerfeld Center, Ludwig-Maximilians-Universit\"at, Theresienstra{\ss}e 37, 80333 M\"unchen, Germany}
%\date{\today}

%%%%%%%%%%%%%%%%%%%%%%%%%%%%%%%%%%%%%%%%%%%%%%%%%%%%%%%%%%%%%%%%%%%%%
%%%% Abstract

\begin{abstract}
We study the structure of galactic halos within a scalar dark matter model, endowed with a repulsive quartic self-interaction, capable of undergoing the superfluid phase transition in high-density regions. We demonstrate that the thermalized cores are prone to fragmentation into superfluid droplets due to the Jeans instability. 
Furthermore, since cores of astrophysical size may be  generated only when most of the particles comprising the halo reside in a highly degenerate phase-space, the well-known bound on the dark matter self-interaction cross section inferred from the collision of clusters needs to be revised, accounting for the enhancement of the interaction rate due to degeneracy. As a result, generation of kpc-size superfluid solitons, within the parameter subspace consistent with the Bullet Cluster bound, requires dark matter particles to be ultra-light.

\end{abstract}
\maketitle

\section{I. Introduction}

The cold dark matter (CDM) paradigm, in which dark matter is incorporated as a non-relativistic fluid composed of collisionless particles, is in spectacular agreement with observations at large scales. However, it presents several challenges when it comes to galactic scale phenomena \cite{Bullock:2017xww}. Among those, the mismatch between the Navarro-Frank-White (NFW) density profile in inner regions of galaxies and clusters obtained from CDM simulations \cite{Navarro:1995iw} and the density profile inferred from observations, known in the literature as 'the core-cusp problem', appears to be a central one and its resolution will most likely affect the status of some of the other puzzles as well. In particular, there have been claims regarding the excess of the dynamical friction predicted by CDM \cite{Debattista:1997bi,Sellwood:2016,Fornaxold,Tremaine}, which is a sensitive function of the dark matter density-profile (a potentially ameliorating discussion within CDM, in some cases, can be found in \cite{Cole,SanchezSalcedo:2006fa}). Furthermore, observations seem to indicate a significant correlation between the gravitational acceleration of baryons and their distribution in disk galaxies \cite{McGaugh:2016leg,Lelli:2017vgz}. This is usually referred to as the mass discrepancy acceleration relation, which is the generalized version of the baryonic Tully-Fisher relation \cite{McGaugh:2011ac,LelliBTFR}, see also \cite{Famaey:2011kh} and references therein. These are well accounted for by Milgrom's empirical law \cite{Milgrom:1983ca,Milgrom:1983pn,Milgrom:1983zz}, which was originally suggested as a modification of gravity. The origin of these observed relations is not yet understood, but seems to be indicative of some form of interaction between baryons and the dark sector. However, it must be pointed out that depending on the mass-range for dark matter particles, the cosmological constraints on such interactions could be significant \cite{Dvorkin:2013cea}. Although it is still unclear whether a purely gravitational interaction present in CDM, without invoking self- or cross-species interactions, is sufficient for generating such correlations, it seems to be an unlikely scenario \cite{Oman:2015xda} (the possibility of reconciliation has been argued \cite{Read:2016}).

These and other related observations have motivated extensions of the standard paradigm, by giving dark matter some additional properties that could be reflected in a more desirable galactic scale behaviour. One of the directions that has attracted much attention recently concerns self-interacting dark matter \cite{Spergel:1999mh}. If strong enough, interactions could affect the density distribution of central regions of the halo significantly, due to a higher number density of  dark matter particles. In particular,  numerical simulations have revealed that the density profile starts deviating from the NFW profile in regions where particles had the chance to interact at least once throughout the lifetime of the halo \cite{Kaplinghat:2015aga}. The main effect of those interactions is the dark matter redistribution, alleviating the central cusp of the NFW profile.  At large distances from the centre of the halo, where dark matter particles have not had a chance to interact yet due to low densities, the density profile is similar to the one in CDM. 

An interesting version of self-interacting dark matter was considered in \cite{Goodman:2000tg,Slepian:2011ev}\footnote{See also \cite{Boehmer:2007um} for the discussion of the core profile and galactic rotation curves within the Bose-Einstein condensate dark matter scenario.} in the form of a repulsive, sub-eV, scalar field exhibiting superfluidity in galaxies, motivated by theoretical simplicity and the capacity for an additional suppression of the dynamical friction. The considered structure of the dark matter halo had a superfluid core enclosed by an isothermal envelope. In that scenario, dark matter was assumed to be in thermal equilibrium which led to the formation of the superfluid core in the central region, where the temperature of the substance is subcritical (due to high densities). The presence of repulsive self-interaction was vital for the stability of the core, as well as for superfluidity, as we will discuss shortly. The envelope, on the other hand, was stabilized by the thermal pressure of dark matter particles, with the junction condition determined in terms of the finite-temperature equation of state. Imposing the Bullet Cluster constraint and the reproduction of the observed features of rotational curves, the model was claimed to be ruled out in \cite{Slepian:2011ev}. See \cite{Sharma:2018ydn} for a more recent discussion of a spherically symmetric density profile for a thermalized halo, derived from an improved finite-temperature equation of state. The core profile and galactic rotation curves within the Bose-Einstein condensate dark matter scenario have been discussed in \cite{Boehmer:2007um} as well. The same scalar dark matter model has been also considered in the context of its gravitational production at the end of inflation \cite{Peebles:1999fz,Peebles:1999se}.

For attractive bosons, which is the case for axions, one lacks positive pressure and consequently the condensate formed at subcritical temperatures is prone to fragmentation \cite{Guth:2014hsa}. The only possibility for sustaining the macroscopic homogeneous core is to consider the case of ultra-light particles. Achieving macroscopic homogeneity in such a way, in the core of the dark matter halo, was first suggested by \cite{Hu:2000ke}, and was coined as Fuzzy Dark Matter. There, the scale of homogeneity is set by the de Broglie wavelength, and the presence of a kpc-size core requires particles lighter than $10^{-21}{\rm eV}$. The detailed analysis of phenomenological implications of such a scenario was performed in \cite{Hui:2016ltb} (see \cite{May:2021wwp} for recent numerical simulations of the structure formation). The idea of substructure formation has been discussed in the context of the superfluid dark matter scenario as well, see \cite{Schiappacasse:2017ham,Alexander:2019qsh} and references therein.

Recently, the idea of dark matter superfluidity was revitalized in \cite{Berezhiani:2015pia,Berezhiani:2015bqa} as a novel mechanism that could be behind the above-mentioned long-range correlations in galactic dynamics. The idea is to have a superfluid dark matter core in galaxies and utilize phonons to mediate an emergent long-range interaction between baryons submerged within this quantum liquid\footnote{The idea of emergent long-range interactions within an ideal Bose-Einstein condensate and the cosmological implications for the scalar dark matter scenario was first considered in \cite{Ferrer:2000hm,Ferrer:2004xj}. Recently, the idea was revisited for superfluids \cite{Berezhiani:2018oxf}. It was demonstrated that superfluidity shortens the range, but still keeping it much longer than the Compton wavelength of dark matter particles. The potential significance of the mechanism for the galactic dynamics, in the context of the ultra-light dark matter scenario (like Fuzzy Dark Matter), was studied as well.}. It was demonstrated in \cite{Berezhiani:2017tth} that in this incarnation of the superfluid dark matter scenario it was more natural to give up global thermal equilibrium and instead to require thermalization within the central neighbourhood of a halo. However, there much freedom was given in specifying the profile by keeping the superfluid pressure not very closely related to the $2\rightarrow 2$ scattering cross section.\footnote{The reason behind this was the assumption that the thermalization could have been governed by two-body interactions (i.e.\ in the disordered phase the $2\rightarrow 2$ scattering was considered to be dominant), while the superfluid pressure was considered to be determined by three-body interactions.}

In this work, we study the structure of galactic halos within the simplest model of dark matter superfluidity combining ideas of \cite{Goodman:2000tg,Guth:2014hsa,Schiappacasse:2017ham} and \cite{Berezhiani:2017tth} together. The additional ingredient added to the pre-existing ideas being the proper application of the Bullet Cluster bound to sub-eV particles, by including the Bose-enhancement factor in the calculation of scattering rates. The result is the inevitable fragmentation of the thermalized core into superfluid islands due to the Jeans instability. It is demonstrated that the extremely light values for the mass are required to give a kpc-size coherence length. 

One of the theoretically simplest interacting dark matter models can be introduced as a massive scalar field, with quartic self-interaction and minimally coupled to gravity, with the following action
\beq
S=\int{d^4 x\sqrt{-g}\left(\frac{1}{16\pi G}R-|\partial \Phi |^2-m^2|\Phi |^2-\frac{\lambda}{2}|\Phi|^4\right)}\,.
\label{lag}
\eeq
Here, we have chosen to work with a complex ($U(1)$-invariant) field, due to manifest particle number conservation. We could have begun with a real scalar, but considering that we are interested in a non-relativistic substance, the net result would have been identical; in this limit, both theories flow to the nonlinear Schr\"odinger's action.

Before diving into the description of the superfluid scenario, let us begin by recapping the Bullet Cluster constraint \cite{Markevitch:2003at} for such a theory. It is usually invoked as the following bound on the scattering cross section for dark matter particles
\beq
\frac{\sigma}{m}\lesssim 1{ \text{ cm}^2}/{\text{g}}\,.
\label{introcross}
\eeq
Strictly speaking, the bound obtained by \cite{Markevitch:2003at} is for the scattering rate, which for the non-degenerate phase-space translates into \eqref{introcross}. Although the latter has been widely applied to various dark matter models, including the sub-eV mass-range, we will argue that it may not be necessarily legitimate. Having said this, let us put this caveat aside for a moment by focussing on $m\gg {\rm eV}$ and translate \eqref{introcross} into a bound on the coupling constant
\beq
\lambda\lesssim \left(\frac{m}{10~{\rm MeV}} \right)^{3/2}\,.
\label{introlambda}
\eeq
In other words, for dark matter heavier than few MeV, the theory has to be strongly coupled in order to violate the Bullet Cluster bound.

For the superfluid scenario one is mostly interested in sub-eV particles. As it is well known, such a candidate must be a non-thermal relict and should be produced via the axion-like vacuum misalignment mechanism. Consequently, it is expected to be in the form of the condensate on cosmological scales. Requiring the equation of state due to interaction-pressure to be the one for a non-relativistic fluid, one arrives at the following bound for the scattering cross section
\beq
\frac{P}{\rho} \Big|_{\text{equality}}=\frac{\lambda\rho\rvert_{\text{equality}}}{8m^4}\ll1\,,\quad \Rightarrow \quad \frac{\sigma}{m}\ll\left(\frac{m}{2\times 10^{-5}\text{ eV}}\right)^5 \frac{\text{cm}^2}{\text{g}}\,.
\label{MR}
\eeq
We have used $\rho\rvert_{\text{equality}}\simeq 0.4 \text{ eV}^4$ as a dark matter density at matter-radiation equality. Notice that if a fraction of dark matter is produced by a mechanism other than the vacuum misalignment, an additional statistical contribution to the pressure would be generated that could tighten the bound (4).  Interestingly, for light-enough particles \eqref{MR} seems to compete with the merger constraint. However, the real Bullet Cluster bound turns out to be even more restrictive than the naively obtained inequality \eqref{introcross}. Due to phase-space degeneracy for sub-eV particles, the interaction rate is boosted by the bosonic enhancement factor. In particular, denoting the average degeneracy factor by $\mathcal{N}$, the improved version of \eqref{introcross} takes the following form for sub-eV dark matter particles of interest
\beq
\frac{\sigma}{m}\mathcal{N}\lesssim 1{ \text{ cm}^2}/{\text{g}}\,, \quad \Rightarrow \quad \frac{\sigma}{m}\lesssim 10^{-2}\left(\frac{m}{\text{eV}}\right)^4\frac{\text{cm}^2}{\text{g}}\,.
\eeq

Moreover, this improved bound results in the exclusion of a significant part of the parameter space, as we are about to show.

The paper is organized as follows. In section II, we overview some of the key properties of the condensate of interacting particles governed by \eqref{lag}. In section III, we study the conditions that lead to a superfluid phase transition in galaxies and clusters. We show that, in order to have kpc-size superfluid cores, we have to consider a scenario with highly degenerate phase-space, implying an ultra-light mass range for dark matter particles. In section IV, we revise the well known bound on dark matter self-interaction cross section inferred from merging clusters accordingly. 
Section V is devoted to the detailed analysis of the full parameter space for a Milky Way-like halo. We summarize the results in section VI.

\section{II. Superfluid Properties}

In a theory of self-interacting bosons, superfluidity may be achieved through Bose-Einstein condensation. Within the theory given by \eqref{lag}, the condensate of $\Phi$ particles can be well described by a homogeneous classical field configuration with finite number density, which spontaneously breaks the U(1) global symmetry of the Lagrangian. As it is well known, any homogeneous fluid is susceptible to the gravitational Jeans instability above a certain length-scale.   For the theory at hand, the low-energy spectrum of excitations around the homogeneous condensate (phonon spectrum) is given by
\beq
\omega_k^2=-4\pi G\rho +c_s^2k^2+\frac{k^4}{4m^2}\,, \qquad c_s^2\equiv\frac{\lambda\rho}{4m^4}\;,
\label{disp}
\eeq
with $G$ standing for the gravitational constant and $\rho$ denoting the superfluid density\footnote{See, e.g., \cite{Berezhiani:2019pzd} for the derivation.}. 
The first term in \eqref{disp} is a tachyonic contribution responsible for the Jeans instability, the second term describes the energy cost for exciting the sound waves and the last one is the kinetic energy of a massive constituent; in other words, the last term indicates that in order to create an excitation, we need to make a massive constituent of the superfluid mobile along the way. It is straightforward to see that, in order to have stable sound waves, $\lambda$ needs to be positive which corresponds to the case of repulsive bosons\footnote{In the opposite case ($\lambda\leq 0$), the only stabilizing contribution to \eqref{disp} would have been the last term.}, which is essential for superfluidity and will be assumed throughout this work.

As one can easily deduce from \eqref{disp}, for the homogeneous condensate, the modes softer than the critical momentum-scale $k_*$ are unstable; with
\beq
k_*^2\equiv2 m^2 c_s^2\left( -1+\sqrt{1+\frac{4\pi G\rho}{m^2c_s^4}} \right)\,.
\label{kstar}
\eeq
The existence of this Jeans scale implies that due to gravity there is an upper bound on the coherence length for the homogeneous superfluid configuration, which is the result of the equilibrium between the gravitational attraction and either the repulsive self-interaction (giving rise to the sound-speed) or the quantum pressure (\`a la Fuzzy Dark Matter scenario).

It is easy to see that, depending on the value of $c_s$, which in turn is determined by the self-interaction strength $\lambda$ (or equivalently by the scattering cross section $\sigma$), the Jeans momentum has two interesting limits depending on the magnitude of the dimensionless quantity
\beq
\xi\equiv \frac{m^2c_s^4}{4\pi G\rho}\,.
\eeq
In particular, in the non-interacting limit $\xi\ll 1 $ we get
\beq
\lim_{\xi\ll  1}k_*^2=\sqrt{16\pi G \rho m^2}\,,
\label{smallcs}
\eeq
which is the Jeans scale stabilized by the quantum pressure. It corresponds to the Fuzzy Dark Matter scenario \cite{Hu:2000ke,Hui:2016ltb} and we refer to it as the \textit{degeneracy pressure case}. For completeness, let us stress that in this limit one cannot talk about sound waves anymore, since the dispersion relation of phonons with wavelength well within the homogeneity domain are highly dominated by the last term of \eqref{disp}. Although subdominant, the presence of the sound-term can still give rise to superfluidity by providing an additional energy cost for excitations. 

For the opposite case, with $\xi\gg 1$, we get 
\beq
\lim_{\xi\gg1}k_*^2=\frac{4\pi G\rho}{c_s^2}\,,
\label{largecs}
\eeq
which is the standard result for the Jeans scale for the superfluid. In this case, the gravitational instability is counteracted and balanced by the repulsive interactions (regular positive pressure). This is the case we are interested in, and we refer to it as the \textit{interaction pressure case.} 

Since the magnitude of the parameter $\xi$ defines the nature of the pressure that sustains the condensate, let us evaluate it at typical galactic densities,
\beq
\xi=\frac{4\pi\mpl^2\rho \sigma}{m^4}\simeq  10^{27} \left(\frac{\sigma/m}{{\rm cm}^2/{\rm g}}\right) \left(\frac{m}{\rm eV} \right)^{-3}\,,
\label{xival}
\eeq
where we estimated $\rho\simeq 10^{-25} {\rm g/cm}^3$, which is the average dark matter density of inner regions of the Milky Way. This is justified as we are after the scenario in which the dark matter density profile is altered significantly only at short scales. In deriving \eqref{xival}, we have used the relation between the sound speed and the scattering cross section
\beq
c_s^2=\frac{\rho}{m^4}\sqrt{2\pi m^2\sigma}\,.
\label{Jscss}
\eeq
As it can be seen from \eqref{xival}, keeping in mind that we do not wish $m$ to be significantly heavier than eV, the only way one could end up with $\xi\ll 1$ would be to take an extremely small scattering cross section (per mass) $\sigma/m$. 

For $\xi\gg 1$, it is straightforward to derive the self-sustained spherical density profile of a zero-temperature superfluid soliton by solving the equation for the hydrostatic equilibrium and Poisson's equation. For the quartic interaction at hand, the superfluid equation of state is $P=\lambda\rho^2/8m^4$, which leads to the following analytic expression for the self-sustained density profile \cite{chandrabook}
\begin{equation}
\rho(r)=\rho_0\frac{\sin\left({2 \pi r/\ell}\right)}{2 \pi r/\ell} ,
\label{eq:profile}
\end{equation}
where $\ell\equiv2\pi /k_*=\sqrt{\frac{\pi\lambda}{4Gm^4}}$ is the Jeans length in the $\xi\gg1$ limit and $\rho_0$ is the central density of the soliton.  Equation (\ref{eq:profile}) shows the equivalence between the Jeans length $\ell$ and the size of the soliton diameter. It must be noted that the density-independence of $\ell$ is tightly connected with the type of self-interaction present for the dark matter field; it would not have been the case for any other form of the potential. It has proven to be convenient to express the size of the soliton in terms of the cross section \cite{Slepian:2011ev}
\beq
\ell=2\pi\left(\frac{8\pi\mpl^4}{m^5}\frac{\sigma}{m}\right)^{1/4}
\simeq 2~{\rm kpc}\left(\frac{\sigma/m}{{\rm cm}^2/{\rm g}}\right)^{1/4} \left(\frac{m}{{\rm meV}} \right)^{-5/4}\,.
\label{JS}
\eeq
This way, one can get an idea of what it takes to have a macroscopic core.

Therefore, if we take a nearly-zero-temperature homogeneous superfluid of $\Phi$s (that would have been stable in the absence of gravity), it will break into superfluid islands of size $\ell$ and the density profile given by \eqref{eq:profile}, similar to the formation of the cluster of stars from a baryonic cloud.

We would like to finish this section by pointing out that the above discussion applies to a zero-temperature superfluid. In the dark matter context, the initial thermalized region has a finite temperature \cite{Slepian:2011ev,Berezhiani:2015bqa}. Therefore, each superfluid soliton will be dressed in an envelope of normal dark matter particles. The transition area is expected to be located where the density drops below the critical one; i.e. the density for which the de Broglie wavelength becomes shorter than the inter-particle separation. Within the superfluid soliton the densities will be of order of the galactic values (maybe somewhat larger, as they would be the result of a collapse), while the transition density can be roughly estimated as $\rho_c\sim m^4v^3$; with $v$ denoting the characteristic galactic velocity determining the temperature. Assuming $v\sim 10^{-3}$, one gets $\rho_c\sim 10^{-28}{\rm g/cm}^3(m/{\rm eV})^4$. In other words, if the dark matter mass is significantly sub-eV (which will be the range of our interest, as we will show), then the envelope begins at the distance from the centre of the soliton where the densities have dropped by more than few orders of magnitude compared to the core. Taking this into account, it is safe to assume that the significant fraction of the dark matter within the thermalized region will be in the form of the superfluid islands. The rest of the matter will be split between the envelopes (gravitationally bound to solitons) and the inter-soliton gas.

\section{III. Thermalization and Superfluid Formation}
Let us investigate what are the conditions that lead to the formation of a superfluid in the galactic medium. 
We would like to begin by mentioning that it is possible to have an effective condensate without a local thermal equilibrium, as long  as the number of particles within the de Broglie volume is large.  In other words, the high degeneracy enables us to describe the quantum state by a homogeneous field configuration, the perturbations around which obey \eqref{disp}. 
Therefore, the coherence length can be estimated as \cite{Guth:2014hsa}
\beq
\ell\simeq{\rm min}\left( 2\pi/k_*, \lambda_{\rm dB} \right)\, ,
\label{islandsize}
\eeq
with $\lambda_{\rm dB}$ denoting the de Broglie wavelength. If equilibrium is not reached, then $\lambda_{\rm dB}$ is determined by  characteristic dark matter velocities obtained from N-body simulations. Using the value of the virial velocity for a typical galactic halo, it is straightforward to verify that (\ref{islandsize}) will always reduce to $\lambda_{\rm dB}$, unless one considers particle masses even lighter than the one for the Fuzzy Dark Matter scenario. Therefore, in practice, one needs to invoke thermalization in order to even hope to get a macroscopic (kpc-size) core for moderately sub-eV particles.

Following what we said, the Bose-Einstein condensation for weakly interacting bosons sets in if two conditions are satisfied:

\begin{itemize}

\item First, the system must reach the equilibrium. This is achieved after particles had sufficient time to interact and reach the (nearly) maximum entropy state, since otherwise the applicability of the Bose-Einstein statistics would be questionable. The time it takes to reach the equilibrium can be estimated as $t_{\rm eq}>t_1$, with $t_1$ denoting the time it takes each particle to scatter at least once. The longer one waits, compared to $t_1$, the more certain one can be for being close to equilibrium. A more precise statement is beyond the scope of this paper.

\item Second, the de Broglie wavelengths of particles must overlap. This corresponds to the system being colder than the critical temperature $T_c\sim n^{2/3}/m$; with $n$ denoting the particle number density. Physically, what happens is that at high temperatures, for which the de~Broglie wavelength is shorter than the inter-particle separation, the gas of weekly interacting particles behaves as a classical system and the Bose-Einstein distribution is well-approximated by the Boltzmann distribution. Below the critical temperature, on the other hand, the latter fails to adequately describe the system, because indistinguishable particles with overlapping wave-packets start to populate the zero momentum state (in compliance with the Bose-Einstein statistics). In fact, for $T\ll T_c$ almost all particles of the gas are in the ground state.
\end{itemize}

Therefore, we expect the dark matter halo to possess few relevant length scales which are not in a one-to-one correspondence: the first one, the thermal radius $R_T$, identifies the region where interactions are efficient enough to allow thermal equilibrium. The second one is the degeneracy radius $R_{\rm deg}$ within which the de Broglie wavelength exceeds the inter-particle separation. The shortest of these identifies the region where the phase transition is expected to take place. The last, but not the least, is $\ell$ connected to the scale below which the condensate is stable.  Understanding the hierarchy between these scales is vital in order to understand whether the fragmentation takes place or not, namely if galaxies present a single superfluid core or a collection of superfluid substructures.

The formation of a dark matter halo is a non-linear process and as such it is challenging (if not impossible) to establish a precise density profile analytically. For purely gravitationally interacting (standard)  dark matter models, $N$-body simulations reveal the more or less universal density distribution, known as the NFW profile
\beq
\rho(r)=\frac{\rho_0}{\frac{r}{r_s}\left(1+\frac{r}{r_s}\right)^2}\, .
\eeq
The characteristic density $\rho_0$ and the scale radius $r_s$  vary from halo to halo. However, according to simulations, there exists a tight relation between these two parameters, known as the mass-concentration relation \cite{Dutton:2014xda}. The size of the halo itself is conventionally defined by the virial radius $R_V$,  which represents the radius within which the average density of the halo (denoted as $\rho_{200}$) is about 200 times the critical density. For our qualitative analysis we begin with the NFW profile and examine under what conditions a significant modification of the profile, followed by a superfluid formation, takes place. As we have already pointed out, one should expect the aforementioned thermalization and degeneracy requirements to be more easily satisfied in central (high density) regions.

Depending on the parameters of the model $m$ and $\lambda$, dark matter particles could become degenerate at densities lower than the ones at which the equilibrium can be reached. In that case, the interaction rate, responsible for thermalization, will be assisted by the degeneracy factor that roughly counts the number of particles in the de~Broglie volume. In general,  the relaxation rate for highly degenerate particles can be estimated as \cite{Sikivie:2009qn,Erken:2011dz}
\beq
\Gamma=\frac{\sigma}{m}\rho v\mathcal{N}\,, \qquad \mathcal{N}={\rm max}\left\{ 1,~  \frac{\rho}{m} \left( \frac{2\pi}{mv} \right)^3\right\}\,,
\label{eq:int}
\eeq
where $v(r)=\sqrt{\frac{G M(r)}{ r}}$ stands for the orbital velocity of dark matter particles and the velocity dispersion has been assumed to be of order $v$, while $M(r)$ is the mass of the halo enclosed in an orbit of radius $r$.  

  Following our earlier discussion, we estimate the thermal radius $R_T$ as the one within which the particles had a chance to scatter at least once throughout the lifetime of the galaxy;  i.e.\ $R_T$ is a radius within which we have $\Gamma t_\text{g}>1$, with $t_\text{g}\approx 13~{\rm Gyrs}$ being the age of the galaxy.

For completeness, let us note that using $t_\text{g}$ as the time-scale for thermalization implies the assumption that it is possible to ignore the phase-space reshuffling of $\Phi$ due to dynamical effects within the galaxy. If this is not justified, the dynamical time $t_{\text{dyn}}=r/v$ is more appropriate to determine the thermal radius.   We will demonstrate in the appendix that the utilization of $t_\text{dyn}$ results in a reduction of the thermalization radius $R_T$ by a factor of few.

Depending on whether the equilibrium is reached while $\mathcal{N}\gg 1$ or not, there could be two qualitatively distinct cases to consider:

\begin{itemize}

\item[{\bf (i)}] If thermalization is reached at radius $R_T$ while $\mathcal{N}\simeq 1$, then we would have a non-degenerate classical gas of weakly interacting particles at distances $r>R_T$ from the centre of the halo. In other words, for $r>R_T>R_{\rm deg}$ particles are not aware of interactions, for $R_T>r>R_{\rm deg}$ particles had the chance to experience interactions and as such the distribution will be more fuzzed out compared to the non-interacting case. Since the particles are non-degenerate in this region, the profile would resemble the profile one obtains in self-interacting dark matter models. At $r<R_{\rm deg}<R_T$, on the other hand, the Bose-Einstein condensation would take place and we would expect to see the presence of superfluid islands of size $\ell$.

\item[{\bf (ii)}] An alternative scenario would be that the high degeneracy is reached at distances larger than the thermal radius ($R_{\rm deg}>R_T$). In this case, the halo would have a simpler structure. In particular, at $r>R_T$ the density profile would be similar to the one for the non-self-interacting dark matter, i.e.\ like NFW, with the possibility of a BEC-like sub-structure at scales shorter than the de Broglie wavelength. Then we would expect the superfluid phase transition directly at $r<R_T$, populating the corresponding volume with the aforementioned superfluid islands of size $\ell$.

\end{itemize}

Now, we are going to study those two scenarios separately. For definiteness, the numerical estimates will be performed for a Milky Way-like galaxy with the total mass $M_{\rm DM}=10^{12}M_\odot$ and the concentration parameter  $c=R_{\rm V}/r_s=6$.

\section*{Case (i): Non-degenerate Thermalization}

Let us focus on the case where the thermalization of  $\Phi$s happens in a non-degenerate setting.  The thermal radius $R_T$ can be estimated as
\beq
\Gamma t_\text{g}=\frac{\sigma}{m}\rho(R_T)v(R_T)t_\text{g}=1\,.
\label{nondegth}
\eeq
For a given density profile, this equality gives $R_T$ as a function of $\sigma/m$. 
Concerning the scaling of \eqref{nondegth} with the parameters of the NFW profile, this is a cumbersome function of $R_T$,  $\rho_0$ and $r_s$. However, the behaviour is simple in limiting cases

\begin{equation}
R_T \simeq { }r_s \Big( \rho_0 r_s \sqrt{2\pi G\rho_0}\frac{\sigma}{m}t_{\rm g} \Big)^{\gamma}\,,\quad {\rm with} \quad \gamma = \begin{cases} {{2}}, & \mbox{for } R_T\ll r_s  \\ 2/7, & \mbox{for } R_T\gg r_s \end{cases}
\end{equation}

Not surprisingly, the overall result for $R_T$ is a monotonically increasing function of the cross section. In other words, for larger $\sigma/m$ the dark matter particles manage to reach equilibrium at larger radii (i.e.\ lower densities). It is easy to find that for the Milky Way-like halo at hand, using $\rho\simeq 10^{-25} {\rm g/cm}^3$ 
, one can conveniently express the thermalization radius as:
\beq
R_T^{\rm MW} \simeq  r_s \left( \frac{\sigma/m}{{\rm cm}^2/{\rm g}} \right)^{\gamma}
\label{mwnondeg}
\eeq

 Let us stress that the case $R_T\ll r_s$ is sensitive to the specific values of $r_s$ and $\rho_0$ and could have strongly been affected by different fits. Moreover, we may see how the strength of the self-interactions affects more mildly $R_T$ in outer regions of the halo: since the density scales as $r^{-3}$, stronger self-interactions are needed to overcome the density drop.

So far, we have not said anything about the mass of the dark matter particle. The equality \eqref{nondegth} will successfully provide us with the estimate for $R_T$, as long as the degeneracy factor is small. Examining the expression for this factor (see \eqref{eq:int}) it is easy to see that to have $\mathcal{N}<1$ at the distance $R_T\ll r_s$ from the centre,
\beq
m\gtrsim 20~ {\rm eV}\cdot \left(\frac{\sigma/m}{{\rm cm}^2/{\rm g}}\right)^{-5/4}\,,
\label{imassbound}
\eeq
implying masses significantly greater than eV. 
 One gets a similar constraint for $R_T\gg r_s$, albeit with a different power law. The important message is that this scenario requires masses above eV, if the interaction strength satisfies the Bullet cluster constraint. This observation together with \eqref{JS}, and using $\sigma/m\lesssim{\rm cm}^2/{\rm g}$, gives us an absolute upper bound on the size of the superfluid soliton
\beq
\ell\lesssim 5\cdot 10^{-2}~ {\rm pc} \,.
\eeq
Interestingly enough, the size of the superfluid patches would be an order of magnitude or so larger than the solar system scale if one were to saturate this, taking the density of the order of the NFW density at our location. However, the density could be somewhat larger as these solitons result from a fragmentation of a locally thermalized dark matter distribution.

\section*{Case (ii): Degenerate Thermalization}
It seems that, as long as  thermalization  
 happens in a non-degenerate setting, particles cannot rely on superfluidity to generate kpc-size solitons. This indicates that we have to explore the possibility of $R_{\rm deg}>R_T$, thus violating \eqref{imassbound}. In this case we have to replace \eqref{nondegth} with its degenerate counterpart
\beq
\Gamma t_\text{g}=\frac{\sigma}{m}\rho(R_T)v(R_T)\mathcal{N}t_\text{g}=1\,, \quad \text{with}\quad \mathcal{N}=\frac{\rho}{m} \left( \frac{2\pi}{mv} \right)^3\gg 1\,.
\label{deg}
\eeq
Here too, the exact expression for $R_T$ is cumbersome. However, similar to the previous case, it can be nicely presented in limiting cases
\begin{equation}
R_T \simeq r_s \left( \frac{4\pi^2\rho_0}{Gm^4r_s^2}\frac{\sigma}{m} t_{\rm g} \right)^{\delta}\,,\quad {\rm with} \quad \delta = \begin{cases} 1/3, & \mbox{for } R_T\ll r_s  \\ 1/5, & \mbox{for } R_T\gg r_s \end{cases}
\label{mwlimits2}
\end{equation}

For a Milky Way-like halo, the expression reduces to
\beq
R_T^{\rm MW} \simeq   \begin{cases} 60\cdot r_s \left[ \frac{\sigma/m}{{\rm cm}^2/{\rm g}}\left( \frac{m}{\rm eV} \right)^{-4} \right]^{1/3}, & \mbox{for } R_T\ll r_s  \\ 10\cdot r_s \left[ \frac{\sigma/m}{{\rm cm}^2/{\rm g}}\left( \frac{m}{\rm eV} \right)^{-4} \right]^{1/5}, & \mbox{for } R_T\gg r_s \end{cases}
\label{mwlimits}
\eeq

Unlike the non-degenerate case, here the size of the thermalized region is determined by $\left({\sigma}/{m}\right)m^{-4}$, as it was previously demonstrated in \cite{Berezhiani:2017tth}. For instance, as one can see from \eqref{mwlimits}, for the galactic halo in question 
\beq
R_T\gtrless r_s\,,\quad \Leftrightarrow \quad  \frac{\sigma/m}{{\rm cm}^2/{\rm g}}\left( \frac{m}{\rm eV} \right)^{-4}\gtrless 10^{-5}\,.
\eeq
Moreover, since \eqref{mwlimits2} depends only mildly on $r_s$ and $\rho_0$, this result is not very sensitive to a fine tuning of the NFW-parameters.

Therefore, it seems there exists a wide range of parameters, for which a significant fraction of the halo has had enough time to have reached equilibrium; the parameter space here is even larger than in case (i), since a specific value of the thermal radius is now generated by different combinations of $m$ and $\sigma/m$ due to the introduction of $\mathcal{N}$.

\section{IV. Bullet Cluster Constraint}
In this section, we would like to revisit the Bullet Cluster constraint \cite{Markevitch:2003at} for the ultra-light bosonic dark matter candidate. The system in question is a merger of  two clusters, in which  the dark matter component is offset with respect to the gas component. As it is well known, the comparison of the observed mass distribution with the simulated one does not seem to indicate the presence of any other dark matter interaction besides the gravitational one. In other words, the average number of scatterings experienced by a given dark matter particle from the bullet cluster, while crossing the target cluster, seems to be less than one
\begin{equation}
\langle n_{sc}\rangle<1 .
\label{eq:number}
\end{equation}
The value $\langle n_{sc}\rangle$ is averaged over all the bullet cluster particles and could be estimated as the product of the interaction rate $\Gamma$ and the crossing time
\begin{equation}
\langle n_{sc}\rangle=\Gamma\frac{2R_V}{v_\text{in-fall}} ,
\label{eq:rate}
\end{equation}
where $v_\text{in-fall}\simeq10^{-2}$ is the in-fall velocity and  $R_\text{V}$ is the virial radius of the target cluster.
Clearly, the bound on the scattering cross section that can be extracted from \eqref{eq:number} and (\ref{eq:rate}) depends on the nature of dark matter.

For example, assuming non-degenerate dark matter that interacts through 2-body interactions
\begin{equation}
\langle n_{sc}\rangle=2\frac{\sigma}{m}R_V\rho .
\label{eq:nondeg}
\end{equation}
Using $\rho\simeq 10^{-25}{\text{ g}}/{\text{cm}^3}$ as an average dark matter density and $R_V\simeq 2$ Mpc, one obtains
\begin{equation}
\frac{\sigma}{m}\lesssim 1{ \text{ cm}^2}/{\text{g}} .
\label{eq:css}
\end{equation} 
The value of the density that we have chosen represents the average density of the Target Cluster within 500 kpc, obtained fitting the matter distribution using an NFW profile \cite{Clowe:2003tk}\footnote{At this distance, the density profile changes from $1/r^3$ to $1/r$: therefore, we are averaging over the region where the mass scales as $r^2$ but not in the part where the mass scales logarithmically with $r$.}.

As we have already discussed from the point of view of dark matter thermalization in galaxies, if particles have a degenerate phase-space, then the interaction rate is enhanced by $\mathcal{N}$ \eqref{eq:int}. The same goes for mergers, if colliding halos are significantly degenerate, then \eqref{eq:css} needs to be reconsidered. It is straightforward to see that for the NFW profile of a cluster (with $R_V\simeq 2~{\rm Mpc}$, $R_V/r_s\simeq 4$ and $\rho_0\simeq 10^{-25}{\text{ g}}/{\text{cm}^3}$) the degeneracy factor exceeds unity everywhere inside the virial radius if $m\ll {\rm eV}$. Notice that $\mathcal{N}_{\text{cluster}}\simeq 10^{-3}\mathcal{N}_{\text{galaxy}}$, since the velocities are an order of magnitude or so higher in clusters. Therefore, if galaxies are strongly degenerate, which is the case for the parameters of interest, then the scattering rates in clusters are expected to be enhanced as well.

We can estimate the improved Bullet Cluster bound on the self-interaction cross section by boosting the interaction rate by a typical (average) value of the degeneracy factor for a cluster, resulting in
\begin{equation}
\label{eq:bound}
\frac{\sigma}{m}\lesssim 10^{-2}\left(\frac{m}{\text{eV}}\right)^4\frac{\text{cm}^2}{\text{g}}\,.
\end{equation}

We would like to stress that this relation applies if the halos are in the form of a gaseous medium of particles with $m\ll {\rm eV}$ and the significant velocity dispersion. If the entire halo of the cluster were to thermalize and undergo fragmentation into superfluid solitons, without significant leftover in the form of a dispersed gas, then \eqref{eq:bound} would need to be ameliorated: in this case, almost all particles would lie in the ground state and transitions to excited states would not be enhanced by degeneracy. However, it seems unlikely not to end up with a significant fraction of particles to remain un-condensed in the process of thermalization and fragmentation. Here, we take this qualitative statement for granted and leave more detailed analysis to future work.

The revised Bullet Cluster constraint (\ref{eq:bound}) leads to the following upper bound on the thermal radius of the Milky Way 
\begin{flalign}
&\left(R_T\right)_{\text{Milky Way}}\lesssim125\text{  kpc} 
 \label{eq:boundTR}
\end{flalign} 
Moreover, it is straightforward to see that the thermal radii of the target and bullet clusters cannot exceed $\approx 0.5 $ Mpc.

Therefore, even if the estimated inequality \eqref{eq:bound} is saturated, only inner regions of the dark matter halo are capable of reaching equilibrium; with the outskirts being unaffected by the presence of self-interactions.
 
\section{V. Relative Size of thermalized region and coherence length}

In this section, we focus on a relation between the Jeans scale $\ell$ and the thermal radius $R_T$. Since we are interested in the superfluid regime of the theory, let us begin by pointing out that for interaction-pressure dominance (large $\xi$) $R_T$ and $\ell$ (given by \eqref{JS}) are controlled by different combinations of $m$ and $\sigma$. Therefore, it may seem possible to pick the parameter values in such a way as to have kpc-size superfluid islands while $R_T$ could vary from values lower than $\ell$ up to $R_V$. We will demonstrate that the dark matter masses required for $\ell\geq R_T$ are so low that one enters the Fuzzy Dark Matter parameter space. This is equivalent to a transition from a superfluid whose degrees of freedom are collective modes (phonons with a linear dispersion relation) to a condensate whose dynamics is described by almost-free constituents (quadratic dispersion relation).

It is straightforward to see that $\ell$, given by \eqref{JS}, can be expressed in terms of $R_T$ using \eqref{deg} as
\begin{equation}
\label{eq:jeans}
\left(\frac{\ell}{2 \text{kpc}}\right)\simeq\left[\mathcal{F}\left({R_T}/{\text{kpc}}\right) \cdot   {\left(\frac{m}{\text{meV}}\right)^{-1}}\right]^{1/4} \ ,
\end{equation}
where $\mathcal{F}$ is determined by the density profile of a halo. The explicit form of $\mathcal{F}$ may be easily deduced from (\ref{deg}). This equation shows that for a given $R_T$ the Jeans scale depends only mildly on the mass of the dark matter particle. Because of this, boosting $\ell$ to sufficiently large values may require lowering $m$ so much that we may end up leaving the interaction-pressure domination regime and enter the quantum pressure dominance. In other words, $\ell$, $\xi$ and $R_T$ depend on three different combinations of $m$ and $\sigma/m$. At the same time, the superfluidity requires $\xi\gg 1$, which is not satisfied by a generic choice of $\ell$ and $R_T$, due to the mild mass-dependence of \eqref{eq:jeans}. 

\begin{figure} 
	\centering
	\includegraphics[scale=0.7]{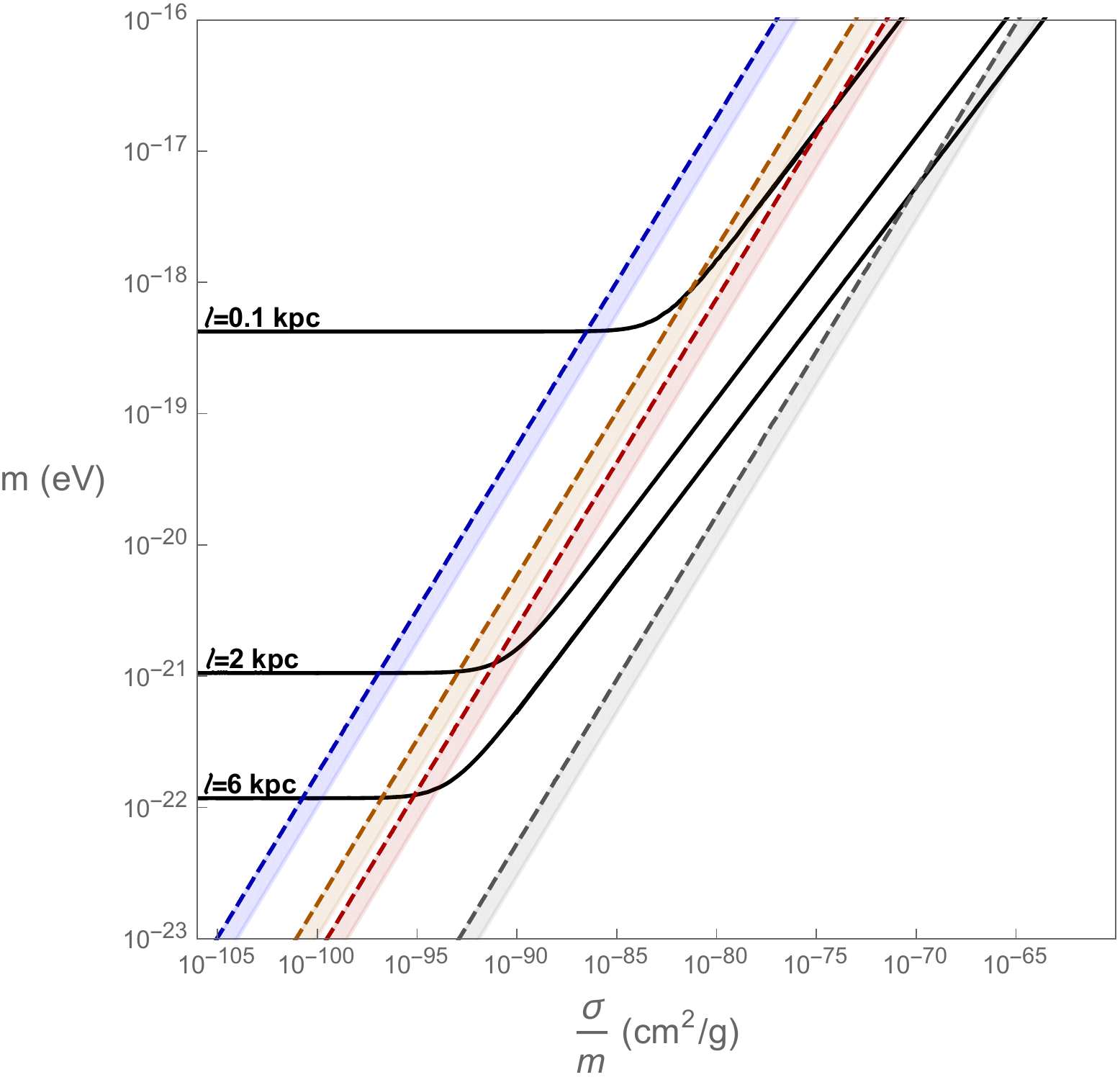} 
	
\caption{Black lines show the slices of  the [m,$\frac{\sigma}{m}$]-parameter space  that generate a Jeans scale of $0.1$ kpc/ $2$ kpc/ $6$ kpc in degenerate regions of the Milky Way dark matter halo, assuming a particle species governed by the Lagrangian (\ref{lag}). They are obtained evaluating \eqref{kstar} numerically. The sloped part of the curve represents the  superfluid regime of the theory ($\xi\gg1$) while the flat part  ($\xi\ll1$) represents the region where    particles self-interact so little as to be considered a non-interacting species. To evaluate $\ell$ we assumed $\rho\simeq 10^{-25} \text{g/cm}^{-3}$, which is the expected average density of the thermal core before fragmentation. This value influences only the $\xi\leq1$ part of the Jeans scale, which is favoured by small $R_T$.
Blue/Orange/Red dashed curves correspond to the parameter space that generates a thermal radius of 0.1 kpc/2 kpc/6 kpc, estimating $R_T$ as the radius within which the particles had the chance to scatter at least once throughout the lifetime of the halo. On the right of the gray line, the Milky Way dark matter halo is in global thermal equilibrium.
In order to stress that we do not really know what is the sufficient number of scattering events for reaching equilibrium, coloured regions show how $R_T$ changes varying the sufficient mean number of scattering events in the interval 1-10.   We see that it is impossible to have an interaction pressure dominated core with $\ell\gtrsim R_T$. }
	\label{fig:general}
\end{figure}

Having made those observations, let us focus on narrowing down the parameter space for which we can get the macroscopic superfluid patches within the thermalization radius. For this we use the expression for the Jeans scale \eqref{kstar} that covers both the superfluid and Fuzzy Dark Matter regimes. The result is given by Fig.1 for a Milky Way-like halo, where different values of the Jeans scale and the thermal radius are shown. Let us stress that, even if $\ell$ is density independent in the interaction pressure case (in case of the $\lambda\Phi^4$-potential), this is not true for $\xi\lesssim1$. Thus, while the sloped part of solid curves is the same in every astrophysical structure, this is not true for the turning point and the flat part.

Now, Fig.1 highlights the impossibility of $\ell\gtrsim R_T$ for halos sustained by the interaction pressure: lines corresponding to a given $\ell$ do not intersect lines describing thermal radii with a smaller relative magnitude in the interaction pressure limit. Therefore, if the dark matter is a scalar particle with a $\lambda\Phi^4$-potential which forms an interaction-pressure-supported superfluid core, we have
\begin{equation}
\ell<R_T \ .
\label{eq:fragm}
\end{equation}

To show that (\ref{eq:fragm}) is not an artefact of our specific definition of the thermal radius, coloured regions represent the parameter space that generates specific values of the  thermal radius when changing the sufficient number of scattering events to equilibrate the dark matter halo in the interval 1-10.

\section{VI. Summary}

Let us conclude by reiterating the qualitative tale of dark matter superfluidity discussed in this work. As dark matter particles begin to clump to form a halo, they start similarly to CDM. Up until the point when particles begin to scatter from each other, they are striving towards an NFW density profile. In regions where densities increase sufficiently for particles to start experiencing collisions, the evolution of the density profile starts to depart from its collisionless counterpart. In fact, the regions within which each dark matter particle has had a chance to scatter few times should be close to equilibrium. Although efficient interactions tend to homogenize the density profile, the phase transition and the formation of superfluid droplets may take place if the de Broglie wavelengths begin to overlap inside thermalized regions. Both equilibration and the overlapping wave-functions favour high densities and are hence easier to achieve within central regions. Denoting the radii of the corresponding boundaries as $R_T$ and $R_{\rm deg}$ respectively, we have demonstrated that a presence of $\sim$kpc-size superfluid patches enforces the parameter space for which $R_{\rm deg}>R_T$. Within $R_T$ the core breaks into superfluid islands of size $\ell$ (determined by the parameters of the model) due to the gravitational Jeans instability.  
The dynamics of the droplets plays an important role to shape the density distribution of the thermal core. As we have shown in the first section, each island is a superfluid soliton (non-topological) with a practically homogeneous core.\footnote{This is similar to the Bose-Einstein condensation of bosonic cold-atoms with attractive contact interactions. As a result, the homogeneous condensate is unstable and fragments into solitons.} For $\ell\ll R_T$, the course-grained (over scales larger than $\ell$) density profile should resemble the one in CDM as superfluid solitons are expected to behave as weakly interacting effective particles.

\begin{figure}[h] 
	\centering
	\includegraphics[height=230pt,valign=t]{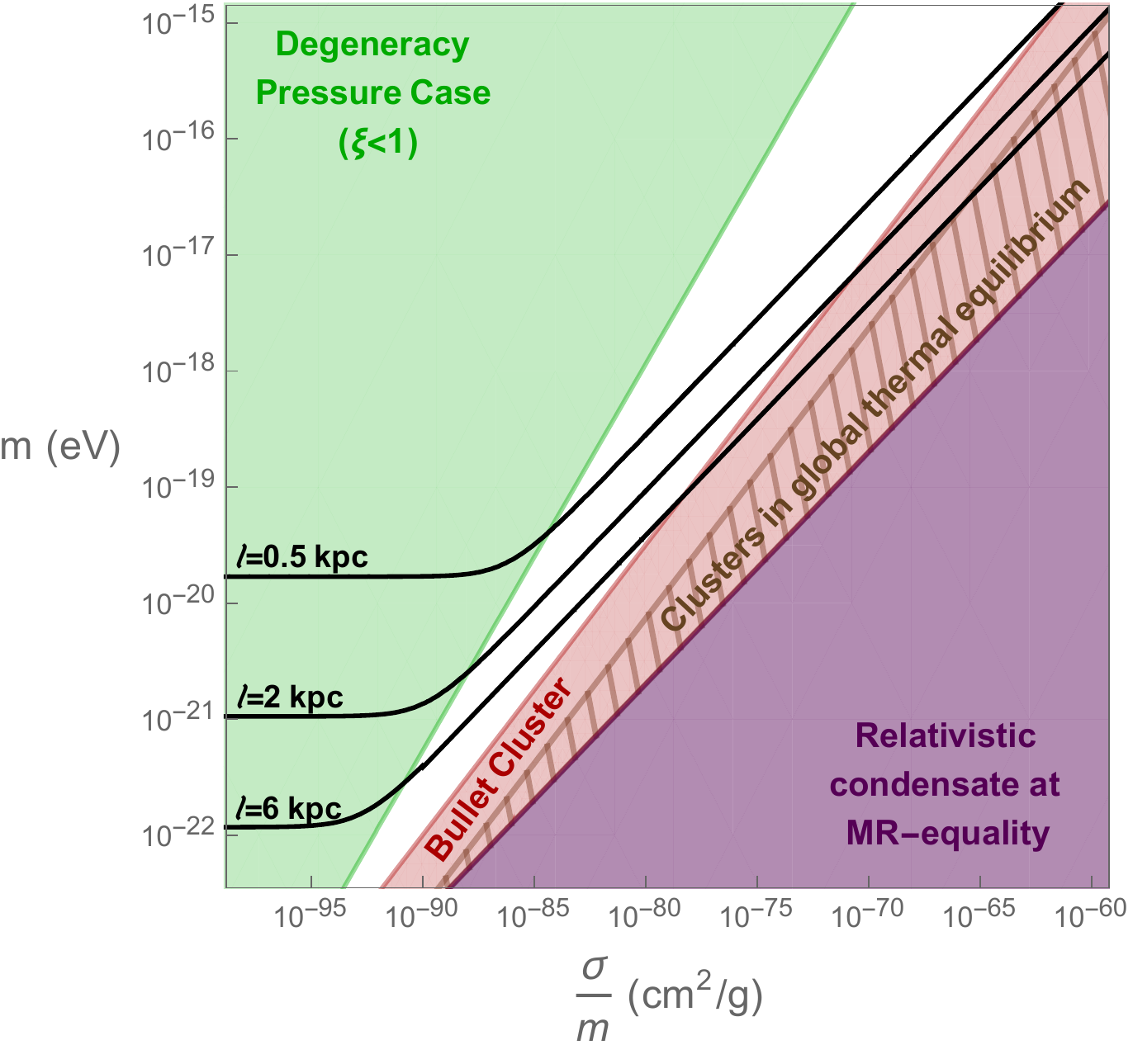}\vspace{0.5cm}
\includegraphics[height=230pt,valign=t]{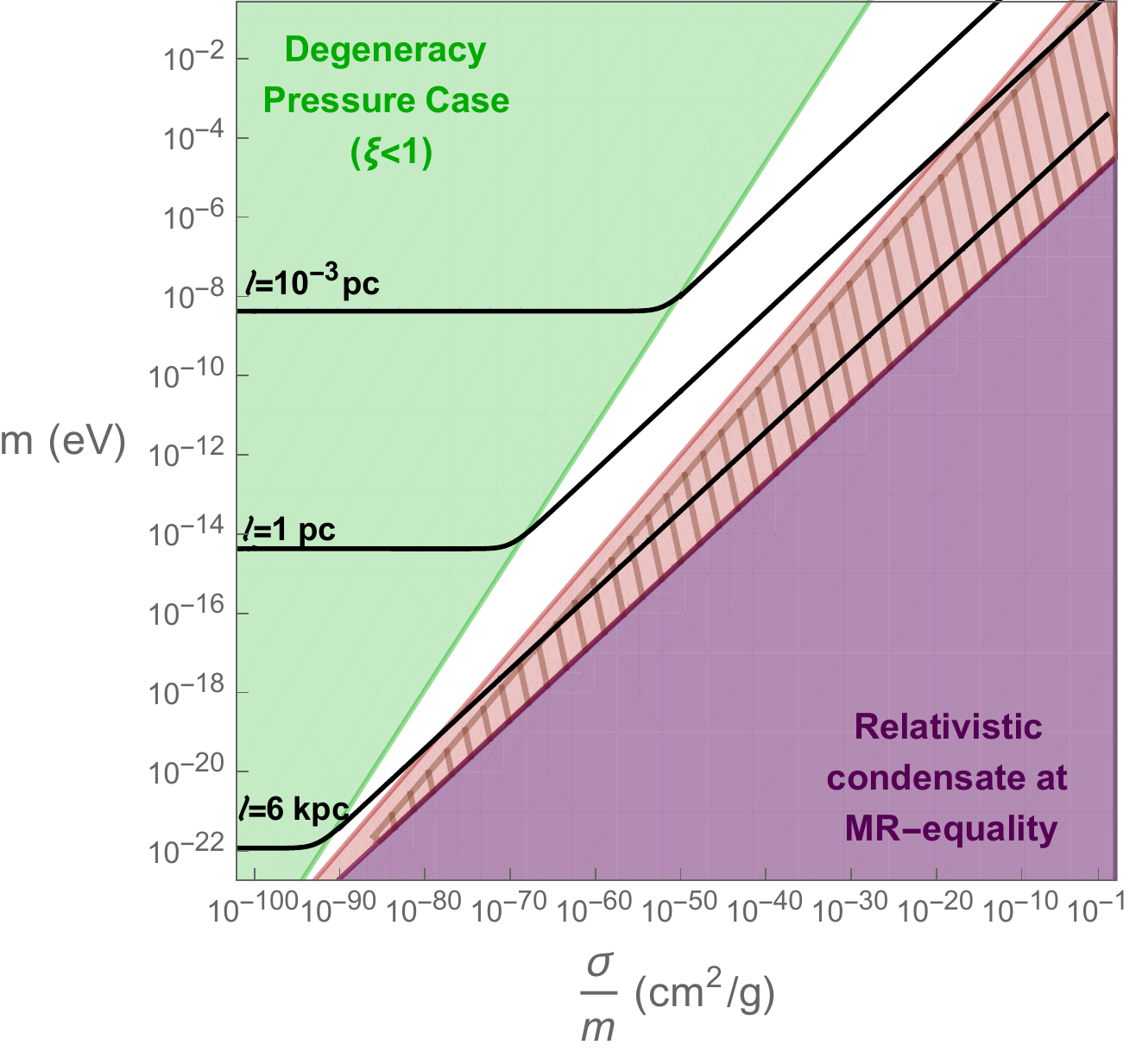}
\caption{Parameter space for $\Phi$. We focussed on the superfluid regime and we excluded halos sustained by the quantum pressure  (green region). The pink region is excluded by the bullet cluster constraint while the purple one represents condensates which were relativistic at matter-radiation equality.  The striped region highlights the parameter space where both the bullet and target cluster are in global thermal equilibrium. In the left panel, Black solid lines identify the parameter space that generates a Jeans scale of 0.5 kpc/2 kpc/6 kpc.  In the right panel, we plotted the whole parameter space, highlighting  $\ell=10^{-3}$~pc/ 1~pc/6~kpc for reference.  }
	\label{fig:fig2}
\end{figure}

In Fig. \ref{fig:fig2} we combine the limits discussed in this work to understand what is the parameter space of a dark matter candidate governed by the Lagrangian (\ref{lag}) that could generate kpc-size solitons in partially thermalized clusters.  
The pink region is excluded by the analysis of the collision of degenerate clusters (leading to an upper limit of the self-interaction strength). For completeness, let us stress that the results of our qualitative analysis are sensitive  to the pre-thermalization shape of the density profile and in particular to the values for the parameters of the NFW distribution. Also, the value we used to estimate the mean density of the thermal core would drop by one order of magnitude if thermalization is strong enough to affect outer regions of the galactic halo. However, this would neither change the behaviour of the theory in the interaction pressure regime, which is density independent, nor the conclusion on the fragmentation of the halo. In fact, the transition point between degeneracy and interaction pressure highlighted by Fig.~\ref{fig:general} would move to the right, since stronger self-interactions are necessary to compensate the smaller mean density of the thermal core.

 Moreover, as we pointed out in Section 4, the bound \eqref{eq:bound} applies only if a significant fraction of the bullet cluster halos is in the gaseous form. This constraint would have been downgraded to \eqref{eq:css} if both the bullet and the target clusters were to thermalize completely and undergo the fragmentation into superfluid solitons, without a significant leftover in the dispersed phase. On the off chance that this could happen, we have highlighted the region of the parameter space in Fig.~\ref{fig:fig2} where both the target and bullet cluster are expected to have thermalized in their entirety. In this case the density profile of clusters would deviate from the usual NFW profile, and the analysis of the profile and the substructure of clusters could lead to constraints on this scenario. The detailed analysis of this scenario being beyond the scope of this work, we have assumed such a fragmentation to result in a significant portion of the cluster mass to have remained in the form of a gaseous medium.

Finally, let us comment on the magnitude of the coupling constant $\lambda$.  Fig. \ref{fig:fig2} illustrates that the generation of solitons with a diameter $\ell\gsim 0.5$ kpc is compatible with the revised bullet cluster bound only for cross sections  $\sigma/m\lesssim10^{-63}\text{ cm}^2/$g and masses $m\lesssim10^{-15}$ eV.   It is straightforward to verify that this means considering $\lambda\lesssim 10^{-65}$:  the smallness of the coupling constant involved is a consequence of the extreme enhancement of the interaction rate by the sub-eV mass of $\Phi$,  both through $\mathcal{N}$ and through the  mass dependence of $\sigma/m$. 

\section*{Achnowledgements}

We would like to thank Justin Khoury for valuable discussions and comments.

\section*{Appendix: Dynamical time as the time-scale for thermalization}  
Let us show the differences emerging by using dynamical time $t_{\text{dyn}}$ to describe the time-scale in which dark matter thermalizes.
 Indeed, if galactic dynamical effects are efficient enough to reshuffle the phase-space of $\Phi$ significantly, the correct time-scale involved would be $t_{\text{dyn}}=r/v$.  Therefore, focussing on the degenerate case, particles had the time to interact only if
 \begin{equation}
\Gamma t_\text{dyn}=\mathcal{N}\frac{\sigma}{m}\rho(R_T)R_T=1 \ .
\label{dyn}
 \end{equation}
 In this case, $\rho_0$, $R_T$ and $r_s$ enter in equation \eqref{dyn} through the combination $\rho^2 R_T/v^3$. Again, we may extract the following limits:
\begin{equation}
R_T \simeq r_s \left( \sqrt{\frac{8\pi^3\rho_0}{G^3 }}\frac{\sigma/m}{m^4r_s^2}\right)^{\delta}\,,\quad {\rm with} \quad \delta = \begin{cases} 2/5, & \mbox{for } R_T\ll r_s  \\ 2/7, & \mbox{for } R_T\gg r_s \end{cases}
\end{equation} 
For a Milky Way-like halo, the expression reduces to
\beq
R_{\rm T, dyn}^{\rm MW} \simeq   \begin{cases} 30\cdot r_s \left[ \frac{\sigma/m}{{\rm cm}^2/{\rm g}}\left( \frac{m}{\rm eV} \right)^{-4} \right]^{2/5}, & \mbox{for } R_T\ll r_s  \\ 10\cdot r_s \left[ \frac{\sigma/m}{{\rm cm}^2/{\rm g}}\left( \frac{m}{\rm eV} \right)^{-4} \right]^{2/7}, & \mbox{for } R_T\gg r_s \end{cases}
\label{RtdynMw}
\eeq

We may now compare \eqref{mwlimits}  and \eqref{RtdynMw}:
\beq
\frac{R_{\rm T, dyn}^{\rm MW}}{R_{\rm T, g}^{\rm MW}} \simeq   \begin{cases} 1/2 \left[ \frac{\sigma/m}{{\rm cm}^2/{\rm g}}\left( \frac{m}{\rm eV} \right)^{-4} \right]^{1/15}, & \mbox{for } R_T\ll r_s  \\ \left[ \frac{\sigma/m}{{\rm cm}^2/{\rm g}}\left( \frac{m}{\rm eV} \right)^{-4} \right]^{3/35}, & \mbox{for } R_T\gg r_s \end{cases}
\label{RtdynMw1}
\eeq

One can see that  $R_T$ gets reduced by a factor 2 at most (the $\sigma$ and $m$ dependence is too mild to affect the ratio by an order one contribution). In this case, the analogue of Fig. \ref{fig:fig2} would reveal a similar picture of what we obtained using $t_{\text{g}}$. 

Finally, we report the upper bound from the bullet cluster on the thermal radius of the Milky Way using $t_{\text{dyn}}$:
\begin{flalign}
&(R^\text{dyn}_T)_{\text{Milky Way}}\lesssim60\text{  kpc} .
\end{flalign}
As expected, the bound on the thermal radius is tighter than \eqref{eq:boundTR}  when we use $t_\text{dyn}$ as the time scale for thermalization.

%\vspace{40pt}

%%%%%%%%%%%%%%%%%%%%%%%%%%%%%%%%%%%%%%%%%%%%%%%%%%%%%%%%%%%%%%%%%%%%%
%%%% formalism
% \newpage

%%%%%%%%%%%%%%%%%%%%%%%%%%%%%%%%%%%%%%%%%%%%%%%%%%%%%%%%%%%%%%%%%%%%%
%%%% Bibliography

%\newpage

\end{document}